\documentclass[10pt,twocolumn,letterpaper]{article}

\usepackage{cvpr}              

\usepackage{graphicx}
\usepackage{amsmath}
\usepackage{amssymb}
\usepackage{booktabs}
\usepackage{multirow}

\usepackage[pagebackref,breaklinks,colorlinks]{hyperref}

\usepackage[capitalize]{cleveref}
\crefname{section}{Sec.}{Secs.}
\Crefname{section}{Section}{Sections}
\Crefname{table}{Table}{Tables}
\crefname{table}{Tab.}{Tabs.}

\def\cvprPaperID{9164} 
\def\confName{CVPR}
\def\confYear{2022}

\begin{document}

\title{Subspace Modeling for Fast and High-sensitivity X-ray Chemical Imaging}

\author{Jizhou Li$^1$, Bin Chen$^2$, Guibin Zan$^1$, Guannan Qian$^1$, Piero Pianetta$^1$, Yijin Liu$^1$\and
$^1$SLAC National Accelerator Laboratory, Stanford University
\and $^2$Max-Planck-Institut für Informatik
}
\maketitle
\begin{figure*}
  \centering
    \includegraphics[scale=0.295]{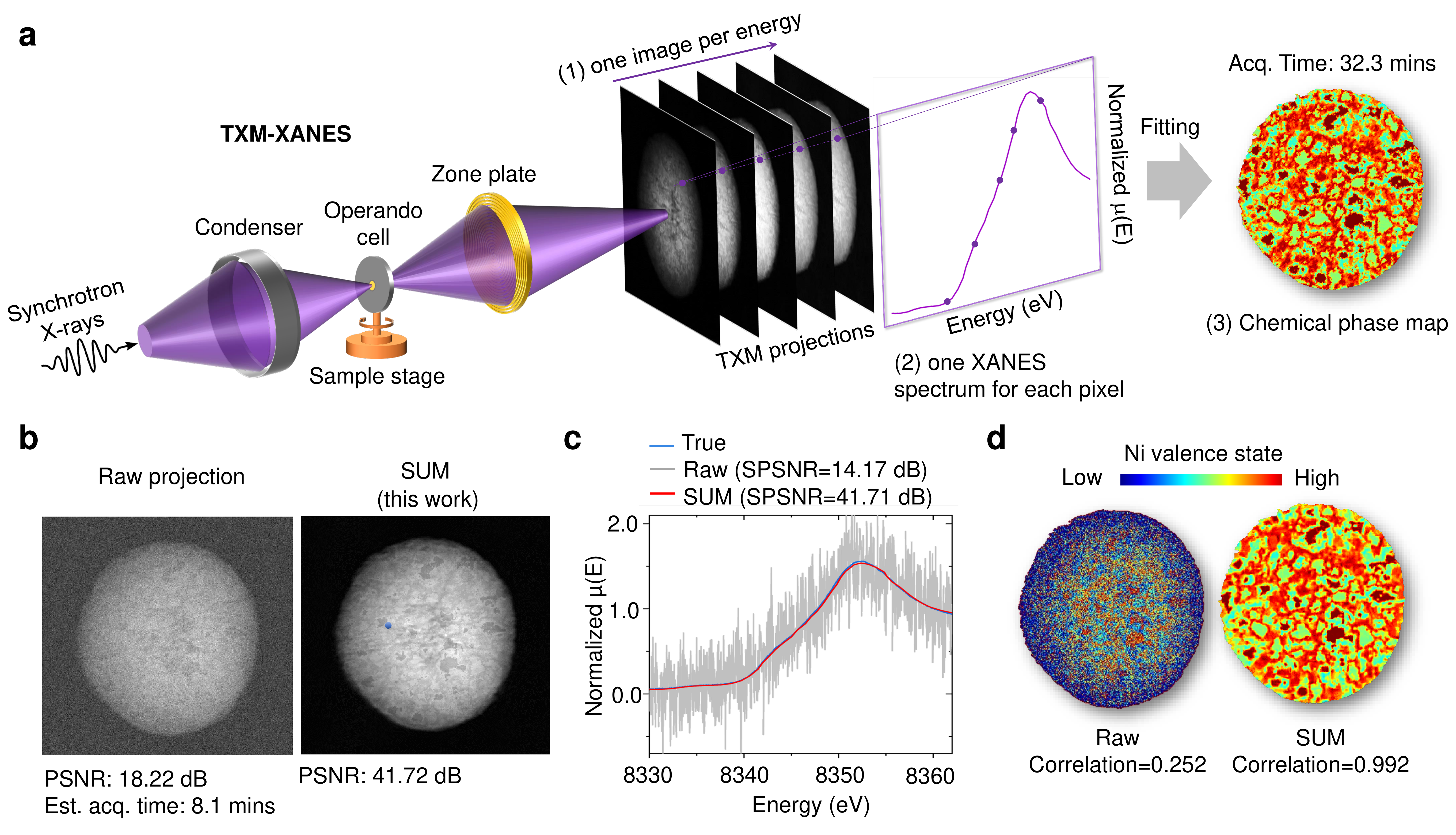}
    \vspace{-6pt}
    \caption{(a) Principles of the TXM-XANES imaging technique. A monochromatic X-ray beam is focused onto the sample using the condenser and the image is projected onto the detector. One image is acquired in absorption contrast at each energy and the XANES spectrum is reconstructed from a series of images with multiple energy points. The normalized XANES spectra are subsequently fitted to create a chemical phase map, representing the chemical states transformation. (b) The proposed approach (SUM) significantly improves the signal-to-noise ratio of the projection images, which enables much faster data acquisition. (c) It also preserves the spectrum information for a single representative pixel, which is indicated as a blue dot in (b). (d) The fitted chemical phase map after denoising has a substantial higher correlation with the ground truth in (a).}
    \vspace{-6pt}
    \label{fig1}
\end{figure*}
\begin{abstract}
  Resolving morphological chemical phase transformations at the nanoscale is of vital importance to many scientific and industrial applications across various disciplines. The TXM-XANES imaging technique, by combining full field transmission X-ray microscopy (TXM) and X-ray absorption near edge structure (XANES), has been an emerging tool which operates by acquiring a series of microscopy images with multi-energy X-rays and fitting to obtain the chemical map. Its capability, however, is limited by the poor signal-to-noise ratios due to the system errors and low exposure illuminations for fast acquisition. In this work, by exploiting the intrinsic properties and subspace modeling of the TXM-XANES imaging data, we introduce a simple and robust denoising approach to improve the image quality, which enables fast and high-sensitivity chemical imaging. Extensive experiments on both synthetic and real datasets demonstrate the superior performance of the proposed method.
\end{abstract}

\section{Introduction}
\label{sec:intro}

X-ray absorption spectroscopy (XAS) is a widely used technique that determines the atomic local structure as well as chemical states in a broad range of disciplines. The conventional XAS lacks spatial resolution which is augmented by the full-field transmission X-ray microscopy (TXM) in recent years~\cite{meirer2011three,wang2014operando,pattammattel2020high}. It offers both high spatial resolution and chemical-specific information through imaging at energy points across the absorption edge of the element of interest. By combining TXM and X-ray absorption near-edge structure (XANES) spectroscopy, two-dimensional and three-dimensional morphological and chemical changes in large volumes can be resolved with sub-50-nm resolution by the TXM-XANES imaging in the hard X-ray regime (5 to 12keV)~\cite{nelson2011three}. An illustration of this technique is shown in~\cref{fig1}a. This technique has been successfully applied to various aspects of materials research, such as the cathode materials’ charge heterogeneity~\cite{wang2014operando,xu2017situ} and mesoscale degradation~\cite{qian2021understanding}.

Despite its wide applicability, TXM-XANES has a disadvantage of relatively slow acquisition process if images at hundreds and thousands of energy points need to be recorded for sufficient energy-resolvable resolution. The fast TXM-XANES imaging is important to robustly resolve the morphological chemical phase transformations, for instance for in-situ/operando battery materials studies. In addition, the long acquisition time may also increase the risk of sample deformation, which in turn makes the data processing and analysis even harder. The speed of TXM-XANES imaging can be accelerated by reducing the energy points or minimizing the X-ray exposure time which is actually more friendly to radiation sensitive samples, in analogy to the low-dose medical X-ray imaging applications~\cite{ravishankar2019image,wang2020deep}. However, the short exposure time eventually yields measurements that are too noisy to be interpreted reliably in the downstream analysis steps. To date, the physical limits of TXM imaging systems are still difficult to be overcome by simply optimizing the microscopy hardware. To alleviate this barrier, computational algorithms could be developed for improving the data quality of short-exposure acquisition, thus pushing the limits of TXM-XANES for fast and high-sensitivity chemical imaging. To the best of our knowledge, this work presents the first attempt that utilized from computational imaging community to address the challenges of X-ray chemical imaging, which is widely available at synchrotron facilities and laboratory-based systems all over the world. 

There have been tremendous efforts over the years in developing better (both traditional and machine-learning based) image restoration algorithms in imaging science. They have been demonstrated to be effective in natural camera~\cite{zhang2021plug,blu2007sure,mao2020image,dabov2007image,Pang_2021_CVPR}, fluorescence microscopy~\cite{luisier2010image,li2017pure,zhang2019poisson,krull2019noise2void,weigert2018content,chen2021three}, electron microscopy~\cite{bepler2020topaz,sheth2021unsupervised}, super-resolution microscopy~\cite{nehme2018deep} and synchrotron X-ray tomography~\cite{liu2020tomogan}, etc. A specific requirement for its application in TXM-XANES imaging is that the algorithm has to be effective and scalable to large datasets. Indeed, in a typical study with TXM-XANES~\cite{qian2021understanding}, two-dimensional XANES images can have image size 1024 $\times$ 1024 pixels (after binning of 2) which cover a field of view of about $16.4 \times 16.4$ um. As reported, data was collected from 288 energy points across the absorption K-edge of Ni from 8100 eV to 9022 eV, which took $\sim$2.5 hours. The acquisition time of three-dimensional measurements is even longer ($\sim$60 hours) resulting in a dataset with size over 500Gb. 

In this work, we aim to explore the intrinsic low-dimensional subspace of TXM-XANES data to improve its image quality from low-exposure acquisition in an unsupervised manner. The key idea of the proposed SUbspace Modeling (SUM) algorithm is that the microscopy images obtained from multiple X-ray energies are highly correlated and all share similar content (sample not moving), which reflects the shape of the target object in the field of view. Their differences are the chemical phases corresponding to illuminations from different energy X-rays. Therefore, it is possible to decompose the noisy data into low-dimensional base and their corresponding coefficients, and then the thresholding can be performed on the coefficients so that the reconstruction becomes noise-free afterwards. This kind of transform-domain approach has been demonstrated to be effective in various image denoising problems~\cite{starck2002curvelet,dabov2007image,blu2007sure}. We use the singular value decomposition (SVD) as the transformer in this work. Importantly, this intuitive idea can be formulated into an optimization framework which follows the ideas of plug-and-play prior~\cite{venkatakrishnan2013plug,wu2020simba} and functional approximation~\cite{blu2007sure,luisier2010image,li2017pure,gilliam2017local}. Its parameters are automatically determined through the statistical modeling of the data and noise, which is implemented as the Stein’s unbiased risk estimator~\cite{blu2007sure,stein1981estimation}. Extensive experiments confirm that the proposed algorithm is able to obtain high quality data from the low-exposure measurements, thus could dramatically reduces the data acquisition time and increases the chemical-resolvability of current TXM-XANES technique.

The main contributions are summarized as follows:

\begin{samepage}
\begin{itemize}
    \item We propose a simple and robust denoising algorithm (SUM) for low-exposure measurements of TXM-XANES data to accelerate the imaging speed and improve the chemical-resolvability.
    \item The SUM algorithm is formulated into an optimization framework. There are no hyperparameters that need to be tuned.
    \item We quantitatively and qualitatively evaluate the proposed algorithm on synthetic and real datasets and show that it performs favorably against other approaches in terms of accuracy and processing speed.
\end{itemize}
  \end{samepage}

\section{Related Work}
\textbf{X-ray Chemical Imaging.} Synchrotrons use electricity to produce intense X-rays that span a broad spectrum. When X-rays interact with an atom or molecule, a variety of signals can result, depending on the type of atom and its chemical environment~\cite{national2006visualizing}. The ability to tune the energy of the incident photons with high resolution opens the vast field of XAS imaging. The applications of X-ray chemical XAS imaging includes dual-energy contrast imaging techniques~\cite{grew2010nondestructive}, in which images directly below and above the absorption edge of a specific chemical element of interest are recorded, as well as the full-field TXM-XANES imaging~\cite{meirer2011three,wang2014operando}, in which a stack of images is obtained at different energies, generating an absorption spectrum (XANES spectrum) for each pixel within the field of view. These spectra are further fit with known reference compounds using a least-square fitting method to determine the edge energy position. The resulting chemical phases for each pixel can be represented as a two-dimensional color map. The concept of TXM-XANES imaging technique is illustrated in~\cref{fig1}. Together with the capability of rotating the sample stage in TXM, the collection of multiple two-dimensional chemical maps at different angles allows tomographic reconstruction with three-dimensional chemical speciation. To handle the system imperfections and speed up the acquisition time, 3D median filtering is often used to improve the signal-to-noise ratio (SNR) of the TXM-XANES imaging data~\cite{meirer2011three,liu2012txm}. Clustering of spectra and then averaging are also proposed to improve the SNR~\cite{qian2021understanding}. The improvement of these approaches, however, is marginal especially for high noise cases. The clustering approach also reduces the spatial resolution of the chemical map, which is suboptimal for high resolution analysis. 

\textbf{Video Denoising.} Denoising has been a long-studied research topic. Numerous works have been proposed in the past~\cite{luisier2010sure,maggioni2012video,liu2010high,buades2016patch}. One of the representative approaches is VBM4D~\cite{maggioni2012video}, that uses motion compensated spatio-temporal patches for video denoising. More recently, impressive results have been demonstrated by data-driven methods~\cite{tassano2020fastdvdnet,ehret2019model,sheth2021unsupervised}. In particular, the unsupervised video denoising methods are mainly based on the Frame2Frame framework~\cite{ehret2019model} where a backbone convolutional neural network pre-trained with supervision is fine-tuned on the video to be processed. However, most of current methods are designed for natural videos. Contrary to the normal settings, TXM-XANES data is more challenging to apply exisiting approaches since spectrum at each pixel spans across hundreds and thousands of frames. It is beneficial to consider all frames for denoising to avoid missing important information. On the other hand, taking account of more frames also allows reducing the exposure time for each single frame thus further accelerating the overall data acquisition process.      

\textbf{Low-Dose X-Ray Imaging.} Much research has been conducted on improving the quality of noisy low-dose X-ray images~\cite{wang2020deep,yang2018low,liu2020tomogan}. For instance, Liu~\etal~\cite{liu2020tomogan} present a denoising technique based on generative adversarial networks to greatly improve the reconstructions of low-dose and noisy synchrotron X-ray data. However, these approaches cannot be directly applied to TXM-XANES data since they are considering a single static image reconstruction (either 2D projection image or 3D tomography). 

\textbf{Plug-and-Play Framework.} Plug-and-play framework that leverages the power of state-of-the-art denoiers has been utilized in various inverse problems~\cite{venkatakrishnan2013plug,chan2016plug,buzzard2018plug,sun2019online,zhang2021plug,ahmad2020plug,yuan2020plug}. It builds on the optimization-based recovery model, where the whole inverse problem is splitted into easier subproblems by handling the forward-model term and the prior term separately, and alternating the solutions to subproblems in an iterative manner. To the best of our knowledge, however, there is no related approach that is well suited for video denoising problem. Part of the reasons for this gap lies in the missing of suitable transformations for specific applications.

\section{Proposed Algorithm}
\label{method}

\textbf{Problem Statement.} Each TXM image is detected by a sensor such as a CCD camera. The times for each X-ray exposure and the energy switching predominantly determine the data quality and overall acquisition time. Similar to visible light, longer exposure time increases photon flux to the image receptor, reduces the noise and subsequently leads to better images with higher signal-to-noise ratios. With the additive noise assumption, the data formation model can be expressed as
\begin{equation}
\label{model}
    \mathrm{\mathbf{y}} = \mathrm{\mathbf{x}} + \mathrm{\mathbf{b}},
\end{equation}
where $\mathrm{\mathbf{x}} \in \mathbb{R}^{MN\times T}$ is the vectorized version of the unknown data. Each column vector is obtained after lexicographic ordering of the TXM image with size $M\times N$ and $T$ is the number of energy points. $\mathrm{\mathbf{y}} \in \mathbb{R}^{MN\times T}$ is the noisy measurement and $\mathrm{\mathbf{b}}\sim\mathcal{N}(0; \sigma^2\mathbf{I})\in \mathbb{R}^{MN\times T}$, $\sigma$ is the standard deviation of noise. The objective of this work is to estimate the underlying true signal $\mathrm{\mathbf{x}}$ from the observed noisy data $\mathrm{\mathbf{y}}$ that is resulting from low exposure time measurement. 

\begin{figure}
  \centering
    \includegraphics[scale=0.21]{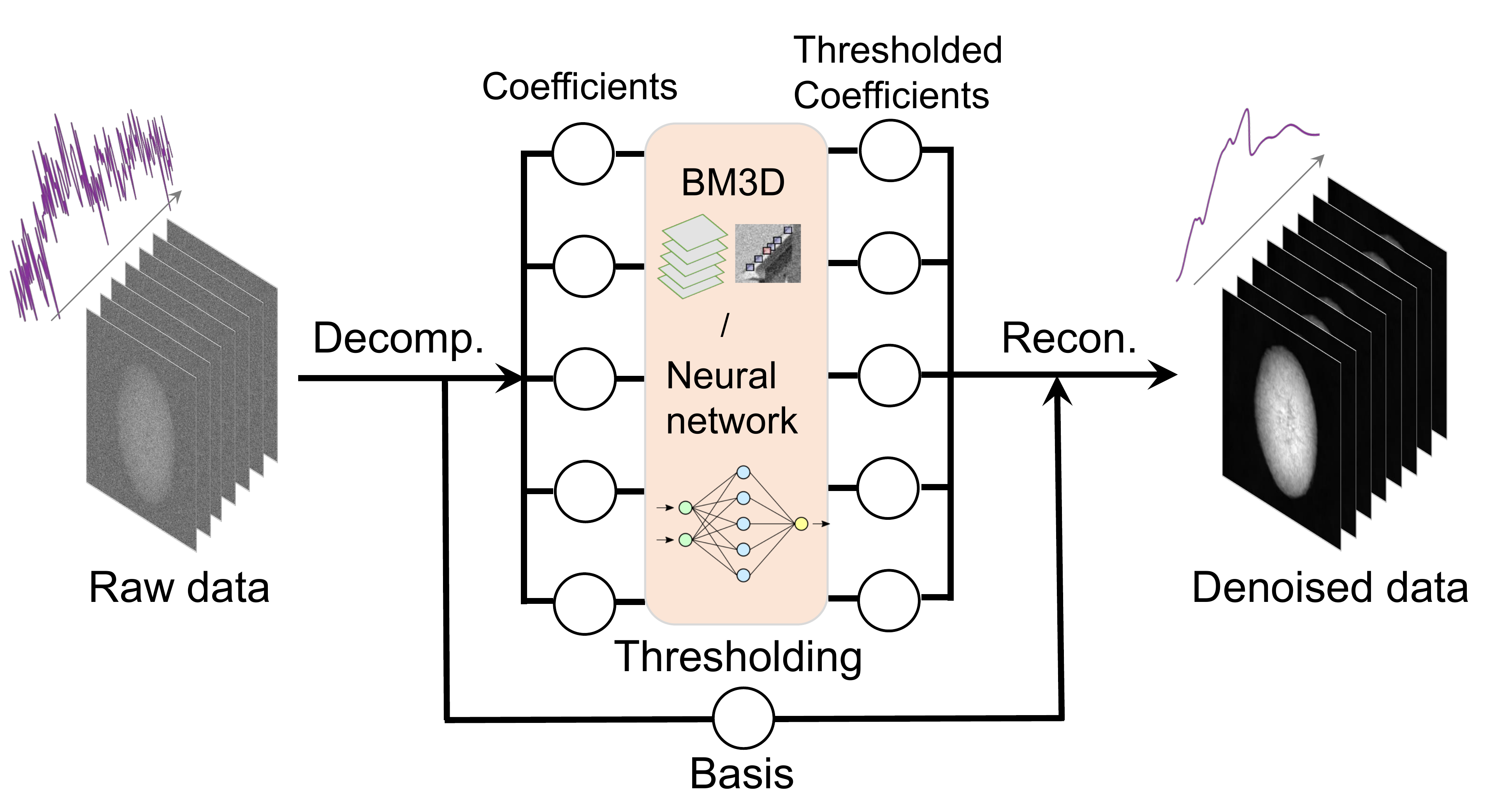}
    \vspace{-10pt}
    \caption{\textbf{Illustration of the proposed algorithm.} The raw noisy data is firstly decomposed by SVD. The coefficients corresponding to different base are thresholded using state-of-the-art image denoising method (\eg BM3D or DnCNN), and then the denoised data is reconstructed.}
    \vspace{-14pt}
    \label{fig2}
\end{figure}
\textbf{The SUM Algorithm.} We introduce a novel algorithm for this task via subspace modeling (SUM). Inspired by the fact that the TXM images with different X-ray energies live in low-dimensional subspace, we can decompose the ground truth data $\mathrm{\mathbf{x}}$ into $K$-dimensional pairs of spatial and ``temporal" components, as $\mathrm{\mathbf{x}}=\mathrm{\mathbf{u}}\mathrm{\mathbf{v}}$, where the columns of $\mathrm{\mathbf{u}} \in \mathbb{R}^{MN\times K}$ can be thought as the basis and $\mathrm{\mathbf{v}}\in \mathbb{R}^{K\times T}$ represents the corresponding coefficients. This decomposition can be simply performed through SVD of $\mathrm{\mathbf{y}}$ under the assumption of the data formation model in~\cref{model}.  

Following the idea of transformation-based denoising methods~\cite{starck2002curvelet,dabov2007image}, particularly the SURE-LET approach~\cite{blu2007sure}, we can fix the basis and try to remove the noise components from the coefficients. The noise-free data can be reconstructed afterwards. That is to say, the denoising problem is formulated as solving the following optimization problem
\vspace{-5pt}
\begin{equation}
\label{reformulation}
    \mathrm{arg}\underset{\mathrm{v}}{\mathrm{min}}\frac{1}{2}\|\mathrm{\mathbf{u}}\mathrm{\mathbf{v}} - \mathrm{\mathbf{y}}\|^2 + \lambda \mathbf{R}(\mathrm{\mathbf{v}}),
\end{equation} where $\mathbf{R}(\mathrm{\mathbf{v}})$ represents some regularization term enforcing prior knowledge of $\mathrm{\mathbf{v}}$. Since $\mathrm{\mathbf{u}}^T\mathrm{\mathbf{u}}=\mathrm{\mathbf{I}}$, \cref{reformulation} is equivalent to 
\vspace{-5pt}
\begin{equation}
\label{reformulation2}
    \mathrm{arg}\underset{\mathrm{v}}{\mathrm{min}}\frac{1}{2}\|\mathrm{\mathbf{v}} - \mathrm{\mathbf{u}}^T\mathrm{\mathbf{y}}\|^2 + \lambda \mathbf{R}(\mathrm{\mathbf{v}}).
\end{equation} 

The regularization term $\mathbf{R}(\mathrm{\mathbf{v}})$ in \cref{reformulation2} may be difficult to be chosen properly, but it can be well approximated by a linear combination of elementary functions $\mathbf{R}_k(\mathrm{\mathbf{v}}^k)$~\cite{blu2007sure,li2017pure}, that is 
\begin{equation}
    \mathbf{R}(\mathrm{\mathbf{v}}) = \sum_{k=1}^K c_k \mathbf{R}_k(\mathrm{\mathbf{v}}^k),
\end{equation} \vspace{-5pt} 

where $\mathrm{\mathbf{v}}^k$ is the $k$-th coefficient vector of the basis $\mathrm{\mathbf{u}}$ and $c_k$ is the corresponding weight which is simply set to be 1 to reduce the degrees of freedom. Thus \cref{reformulation2} can be solved by decoupling each $\mathrm{\mathbf{v}}^k$ and its solution has a general form as
\begin{equation}
    \label{solution}
    \mathrm{\mathbf{\hat{v}}} = [\mathrm{\mathbf{\hat{v}}}^1; ... \mathrm{\mathbf{\hat{v}}}^k; ... \mathrm{\mathbf{\hat{v}}}^K], \mathrm{\mathbf{\hat{v}}}^k = \mathcal{D}_{\lambda}^k\left(\mathrm{\mathbf{u}}_k^T\mathrm{\mathbf{y}}\right),
\end{equation} and $\mathcal{D}_{\lambda}^k \left(\mathrm{\mathbf{w}}\right)$ is the coefficient thresholding operator, which is defined as 
\vspace{-5pt}
    \begin{equation}
    \label{finalproblem}
        \mathcal{D}_{\lambda}^k \left(\mathrm{\mathbf{w}}\right)= \mathrm{arg}\underset{\mathrm{v}}{\mathrm{min}}\frac{1}{2}\|\mathrm{\mathbf{v}} - \mathrm{\mathbf{w}}\|^2 + \lambda \mathbf{R}_k(\mathrm{\mathbf{v}}).
    \end{equation}
    \vspace{-13pt}

\Cref{finalproblem} can be then regarded as a typical image denoising step, and state-of-the-art image denoising algorithms for additive Gaussian noise, either traditional BM3D~\cite{dabov2007image} or convolutional neural network algorithm~\cite{zhang2017beyond}, can be directly applied here. Relevant discussions on which kind of method works best for the TXM-XANES data in~\cref{analysis} and we chose DnCNN~\cite{zhang2017beyond} for the experiments in this work. The final output can be then reconstructed by $\mathrm{\mathbf{\hat{x}}} = \mathrm{\mathbf{u}}\mathrm{\mathbf{\hat{v}}}$. An illustration of the proposed algorithm is shown in~\cref{fig2}. 
\begin{figure}
  \centering
    \includegraphics[scale=0.21]{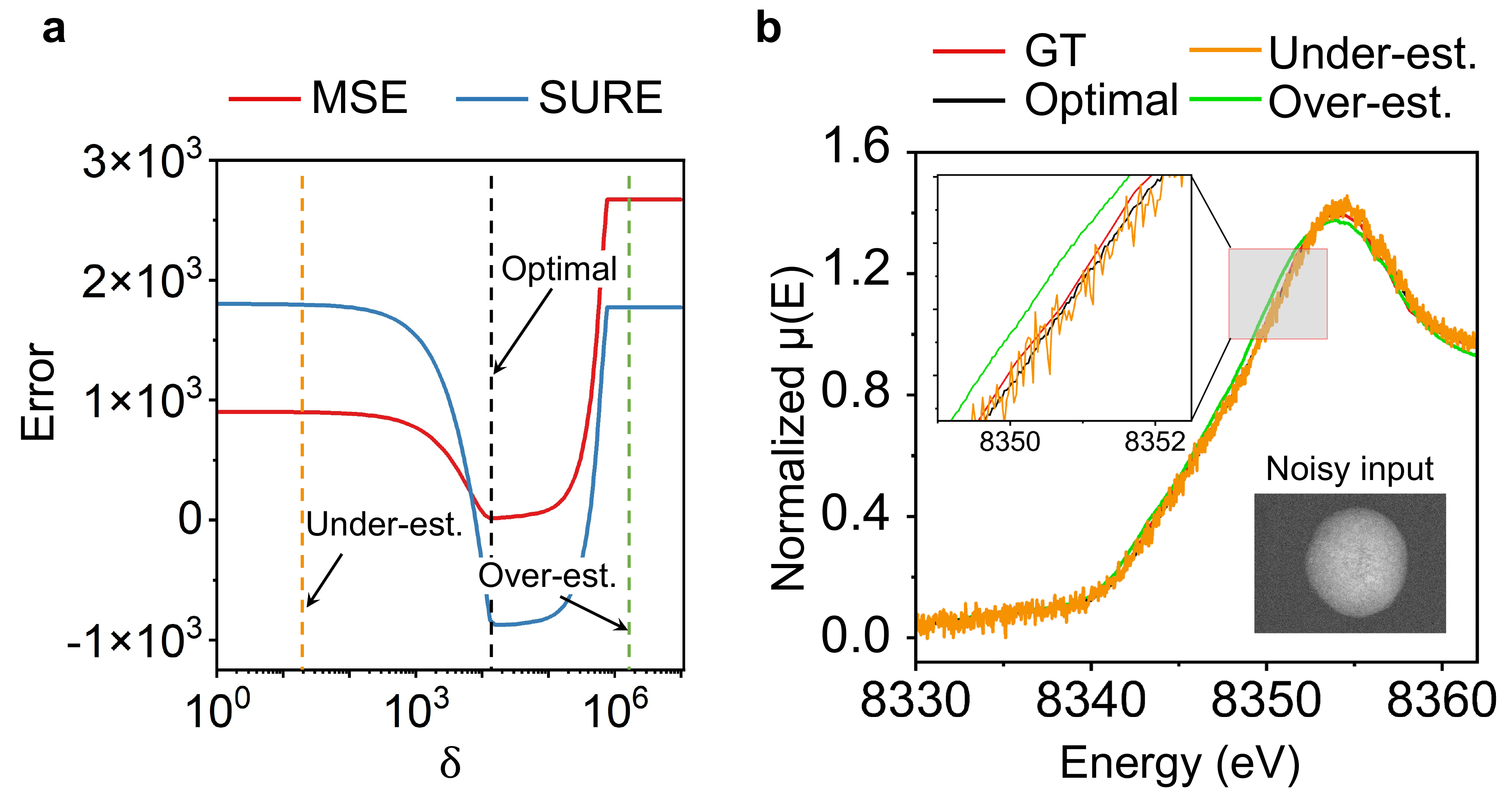}
    \vspace{-10pt}
    \caption{\textbf{Automatic hyperparameter determination.} (a) The optimal thresholding singular value ($\delta$) is determined by minimizing the mean squared error (MSE) and its unbiased estimate (SURE). (b) Demonstration of the effect of different $\delta$'s in (a) on the obtained spectrum by the proposed denoising algorithm. Over-estimated $\delta$ leads to over-smoothed spectrum, while under-estimated $\delta$ gives noisy spectrum.}
    \vspace{-12pt}
    \label{fig3}
\end{figure} 

The central idea of this approach is from the typical transformation-based image restoration method, where SVD serves as the transformer and the inner image denoising algorithm is used to threshold the coefficients. Note that the noise associated to the coefficients is still following Gaussian distribution with the same variance as $\mathrm{\mathbf{y}}$. On the other hand, the overall algorithmic workflow is in align with the recent trend of plug-and-play image restoration methods~\cite{venkatakrishnan2013plug,wu2020simba}, in which the off-the-shelf Gaussian denoiser is utilized to solve various inverse problems instead of using hand-crafted image priors in the optimization. Most notably, instead of dealing with a high-dimensional video data, the dimension of the restoration problem is effectively reduced to 2D image denoising problem by the proposed approach.

\textbf{SURE-based Parameter Selection.} The proposed algorithm is simple and effective to improve the quality of TXM-XANES imaging data. However, the size of the low-dimensional subspace $K$ is a critical parameter to achieve the optimal performance. Too many components result in poor representations while too few lead to large variances. To find the correct trade-off $K$ for the proposed algorithm $\mathrm{\mathbf{\hat{x}}} = \mathcal{F}_K(\mathrm{\mathbf{{y}}})$, we revisit that our essential goal is to minimize the mean squared error (MSE) between the estimate $\mathrm{\mathbf{\hat{x}}}$ and ground truth $\mathrm{\mathbf{{x}}}$, that is to minimize 
$
\label{MSE}
    \mathrm{MSE} (\mathrm{\mathbf{\hat{x}}},\mathrm{\mathbf{{x}}}) = \frac{1}{MNT}\mathbb{E}\left\{\|\mathcal{F}_K (\mathrm{\mathbf{y}}) - \mathrm{\mathbf{x}}\|^2\right\}.
$ In practice, $\mathrm{\mathbf{x}}$ is unknown but \text{MSE} can be well approximated by the Stein’s unbiased risk estimate (SURE)~\cite{stein1981estimation}, which is only related to the observation $\mathrm{\mathbf{y}}$. Thus instead of optimizing the MSE, in which the ground truth is unknown, we can directly minimize the SURE to determine the optimal parameter. The formula and derivation procedure of SURE is similar to those of other works~\cite{blu2007sure,li2017pure,luisier2010sure}, thus omit here. As shown in~\cref{fig3}, the optimal thresholding values obtained by MSE and SURE minimization are identical. 

\begin{figure}
  \centering
    \includegraphics[scale=0.32]{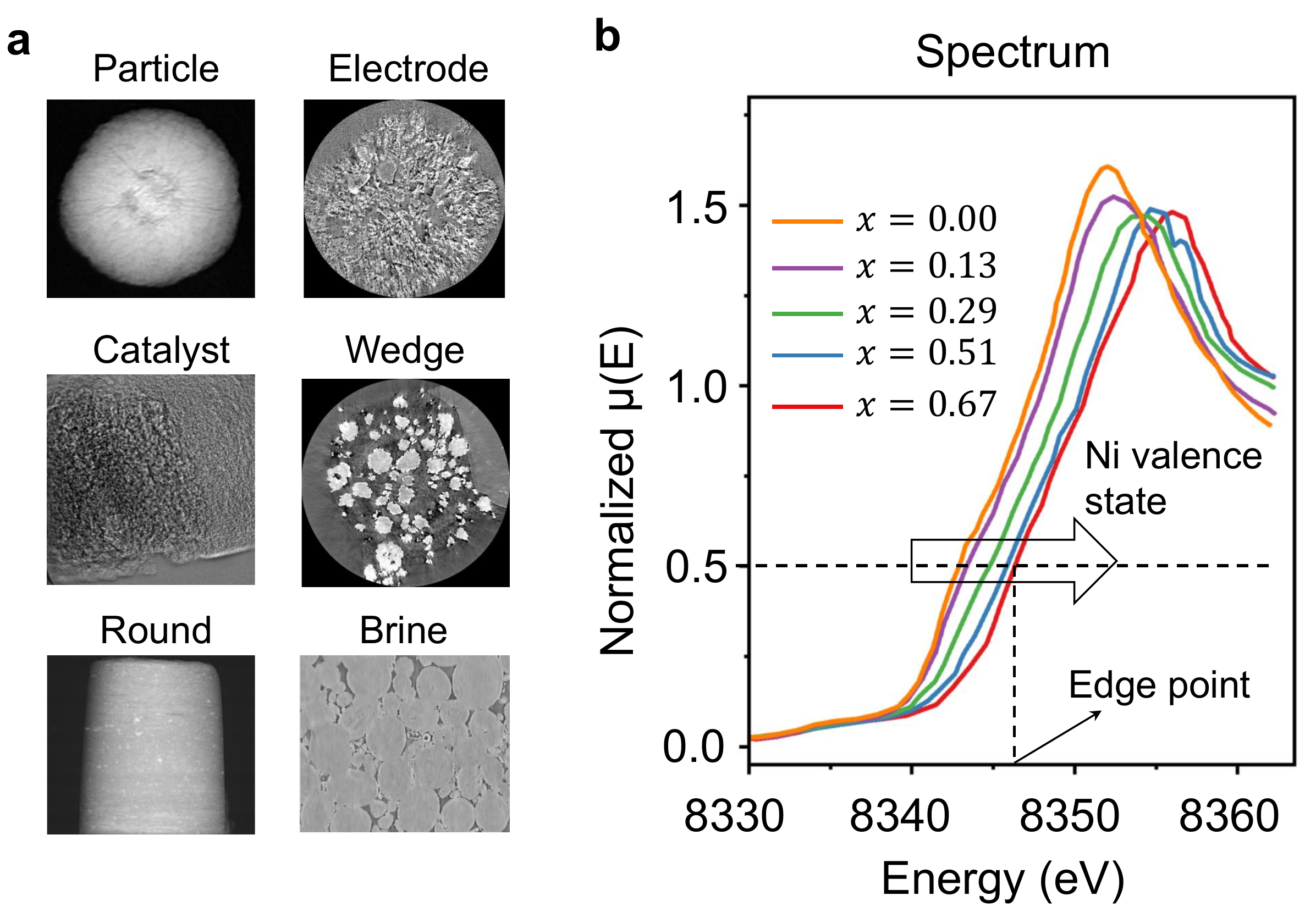}
    \vspace{-6pt}
    \caption{\textbf{Experimental settings.} (a) Typical examples of the test datasets: projections (left) and reconstructed slices (right). (b) Normalized spectra under different Ni valence states of hard X-ray XANES in a battery cathode $\mathrm{Ni}_x\mathrm{Co}_y\mathrm{Mn}_{1\text{-}x\text{-}y}\mathrm{O}_2$. $x$ and $y=0.1$ represent the percentage of elements Ni and Co in the sample, respectively.}
    \vspace{-15pt}
    \label{fig6}
\end{figure} 
\vspace{-5pt}
\section{Experiments and Results}
\subsection{Experimental Settings and Evaluation Metrics}
We perform experiments over six different datasets, as shown in~\cref{fig6}a, which consist of three X-ray projection images (\textit{Particle}, \textit{Catalyst} and \textit{Round}) and three reconstructed slices (\textit{Electrode}, \textit{Wedge} and \textit{Brine}). Image details can be found in \textbf{Suppl. A}. They are used to simulate the scenarios of 2D and 3D TXM-XANES imaging, respectively. In total, there are 642 images for the testing. The simulated movie data is generated by randomly assigning five spectra (\cref{fig6}b) to the pixels in the image. These spectra correspond to different Ni valence states of hard X-ray XANES in a battery cathode $\mathrm{Ni}_x\mathrm{Co}_y\mathrm{Mn}_{1\text{-}x\text{-}y}\mathrm{O}_2$ (NCM). $x$ and $y$ represent the percentage of elements Ni and Co in the sample, respectively. The edge point (energy at 0.5 spectrum position) is an important parameter to probe the chemical state transformation during battery cycling. 

Each movie is further corrupted with additive Gaussion noise at different noise levels $\sigma\in [10, 150]$. The performance of the denoising algorithms is assessed in terms of three criteria:
\vspace{-2pt}
\begin{itemize}
  \setlength\itemsep{0.2em}
    \item \textbf{FPSNR}: Frame peak-signal-to-noise ratio (PSNR), which is calculated by the average of the PSNRs of each frame.
    \item \textbf{SPSNR}: Spectrum PSNR, which is the average of the PSNRs of each spectrum in the image pixels.
    \item \textbf{Correlation}: The correlation coefficient between the obtained and ground truth chemical maps. 
\end{itemize}

\begin{figure}
  \centering
    \includegraphics[scale=0.18]{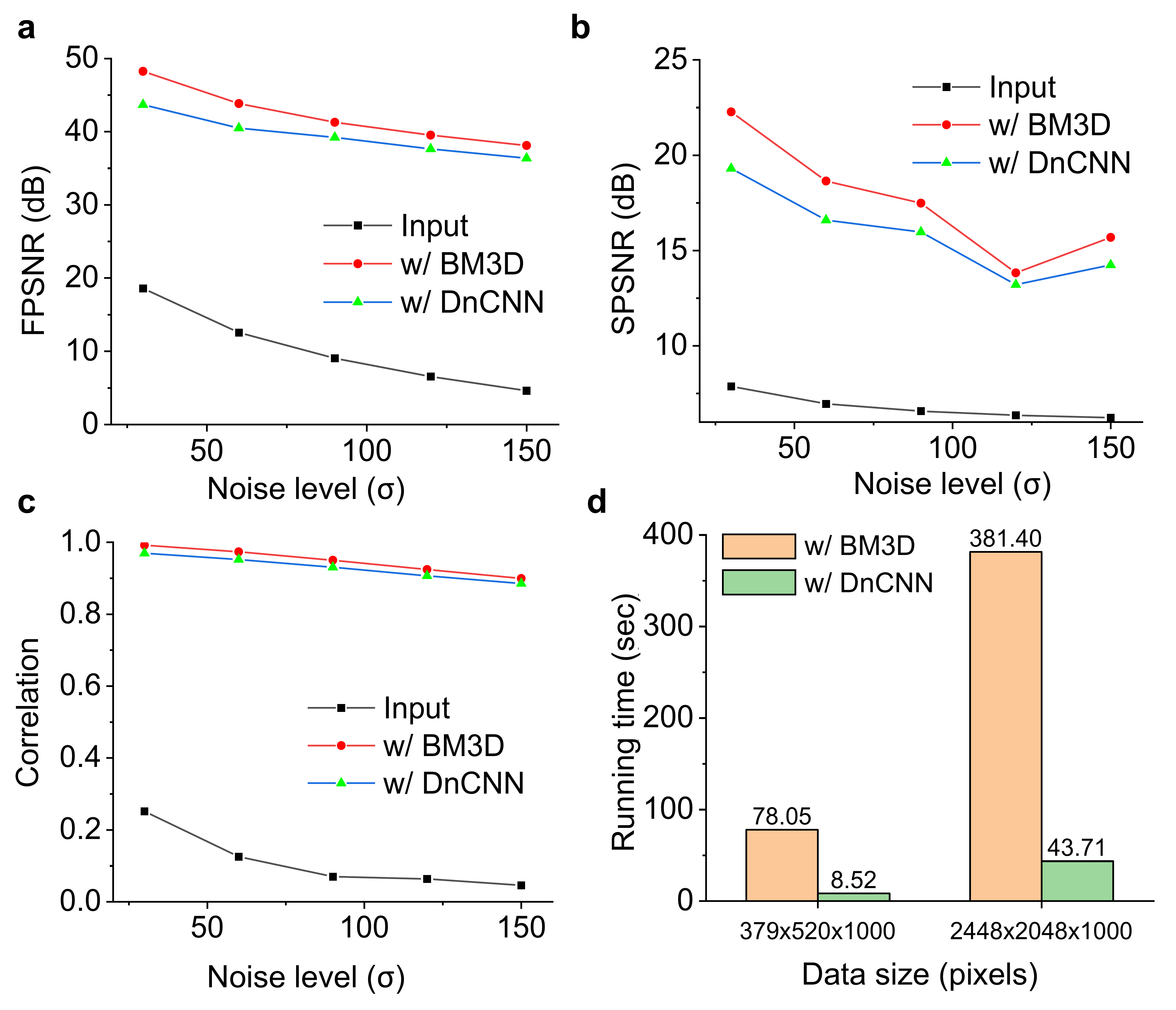}
    \vspace{-18pt}
    \caption{\textbf{Selection of the off-the-shelf denoiser in SUM.} Two typical algorithms (BM3D and DnCNN) are evaluated in terms of frame PSNR (a), spectrum PSNR (b), correlation to chemical map (c) and running time (d).}
    \vspace{-10pt}
    \label{fig4}
\end{figure} 

All experiments are carried out on a workstation with Intel Core i9 and Nvidia Titan V GPU. 
\begin{figure}
  \centering
    \includegraphics[scale=0.18]{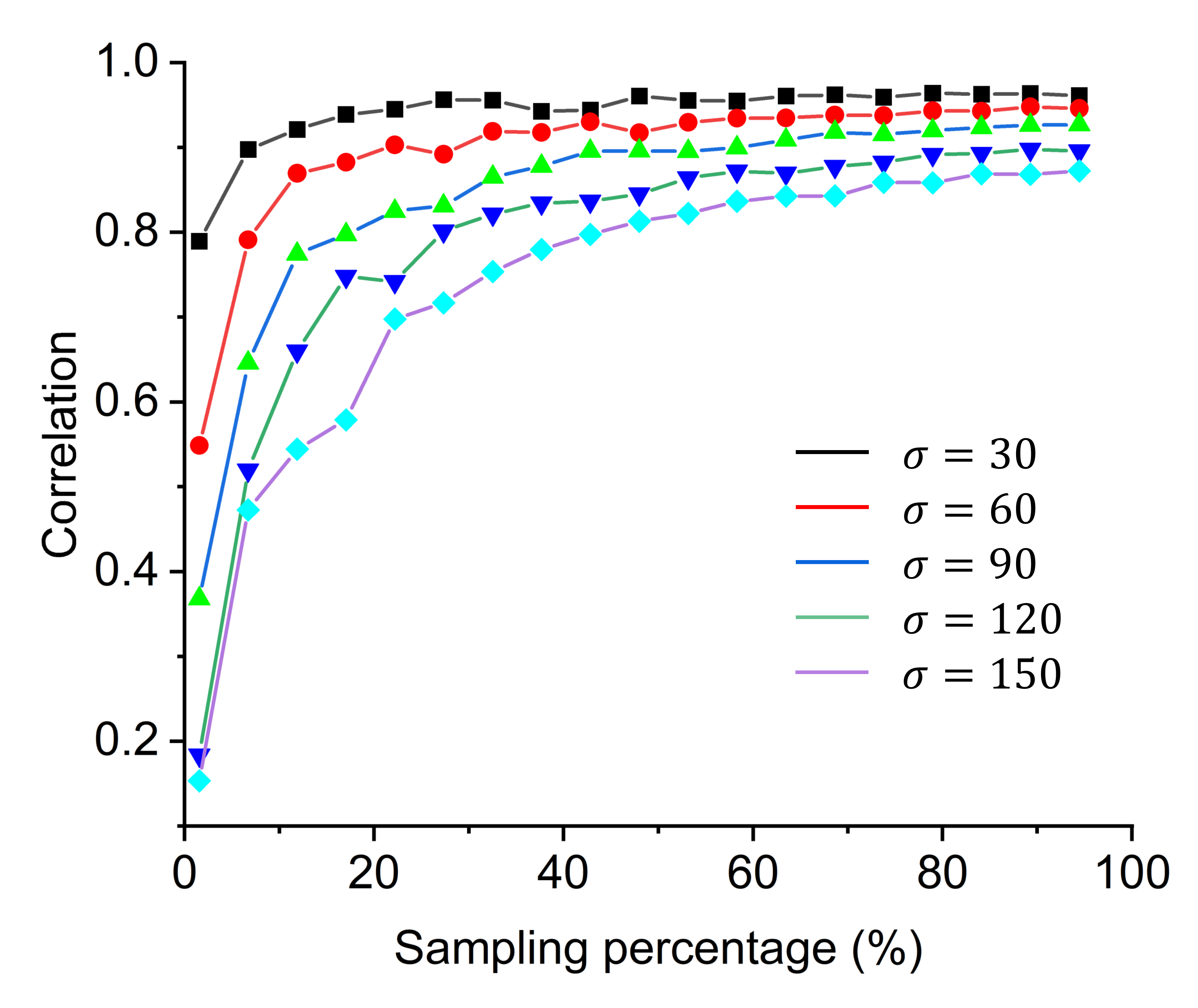}
    \vspace{-10pt}
    \caption{\textbf{Denoising helps reduce the sampling percentage.} The correlation of the chemical maps obtained from the proposed algorithm SUM and the ground truth with respect to different sampling percentages. SUM has the potential to reduce the number of energy points.}
    \vspace{-10pt}
    \label{fig5}
\end{figure} 

\begin{table*}[ht]
    \centering
    \footnotesize{
    \begin{tabular}{cccccccccc}
        \toprule

        \multicolumn{1}{l}{\phantom} &
        \multicolumn{1}{l}{\phantom} &
        \multicolumn{1}{l}{\phantom} &
        \multicolumn{2}{c}{Baseline}    &
        \multicolumn{2}{c}{Traditional}    &
        \multicolumn{2}{c}{Unsupervised CNN}  &
        \multicolumn{1}{c}{This work} \\ 
        \cmidrule(lr){4-5}
        \cmidrule(lr){6-7}
        \cmidrule(lr){8-9}
        \cmidrule(lr){10-10}
        
         \multicolumn{1}{c}{Test set} &
        \multicolumn{1}{c}{$\sigma$} &
        \multicolumn{1}{c}{Noisy} &
         \multicolumn{1}{c}{MedFilt3} &
         \multicolumn{1}{c}{SVD} &
        \multicolumn{1}{c}{ReLD} &
        \multicolumn{1}{c}{VBM4D}     &
        \multicolumn{1}{c}{UDVD (5 frames)} & 
         \multicolumn{1}{c}{RFR}     &
        \multicolumn{1}{c}{SUM}\\

        \midrule
        
        \multirow{3}{*}{Particle} 
        & $10$ & 28.13 / 0.76 & 35.19 / 0.85 &  \textbf{46.69 / 0.99} & 43.88 / \textbf{0.99}  & 43.52 / 0.97 &   32.90 / 0.75 & 18.07 / 0.40 & 43.19 / 0.98 \\
        & $60$ & 12.57 / 0.14 & 24.60 / 0.31 &  29.98 /0.84 & 29.81 / 0.63 & 32.14 / 0.51 & 21.73 / 0.40 & 16.94 / 0.17 & \textbf{39.07 / 0.96} \\
        & $150$ & 4.61 / 0.05 & 16.91 / 0.11 & 23.64 / 0.45 & 21.03 / 0.14 & 28.08 / 0.21 &  14.08 / 0.12 & 15.14 / 0.05 & \textbf{35.23 / 0.91} \\
        \midrule
        \multirow{3}{*}{Wedge} 
        & $10$ & 28.13 / 0.64 & 33.32 / 0.86 & \textbf{46.80 / 0.99}   & 43.68 / 0.98  & 40.47 / 0.95  & 20.68 / 0.69 &  16.23 / 0.40  & 44.76 / \textbf{0.99} \\
        & $60$ & 12.57 / 0.10 & 24.33 / 0.21 & 30.75 / 0.76  &  29.71 / 0.54  & 29.54 / 0.56  & 18.07 / 0.34 & 15.24 / 0.07 & \textbf{39.31 / 0.96} \\
        & $150$ & 4.61 / 0.03 & 16.89 / 0.08 & 21.82 / 0.30  &   20.95 / 0.12 & 25.35 / 0.23 & 14.84 / 0.23  & 13.97 / 0.03& \textbf{34.40 / 0.89} \\

        \bottomrule
    \end{tabular}
    }
    \vspace{-5pt}
    \caption{\textbf{Denoising results of different approaches}. Performance values are averaged FPSNR (in dB) and the correlation coefficient with the ground truth chemical map, respectively. }
    \label{tab:test_data}
    \vspace{-5pt}
\end{table*}

\begin{figure*}
  \centering
    \includegraphics[scale=0.26]{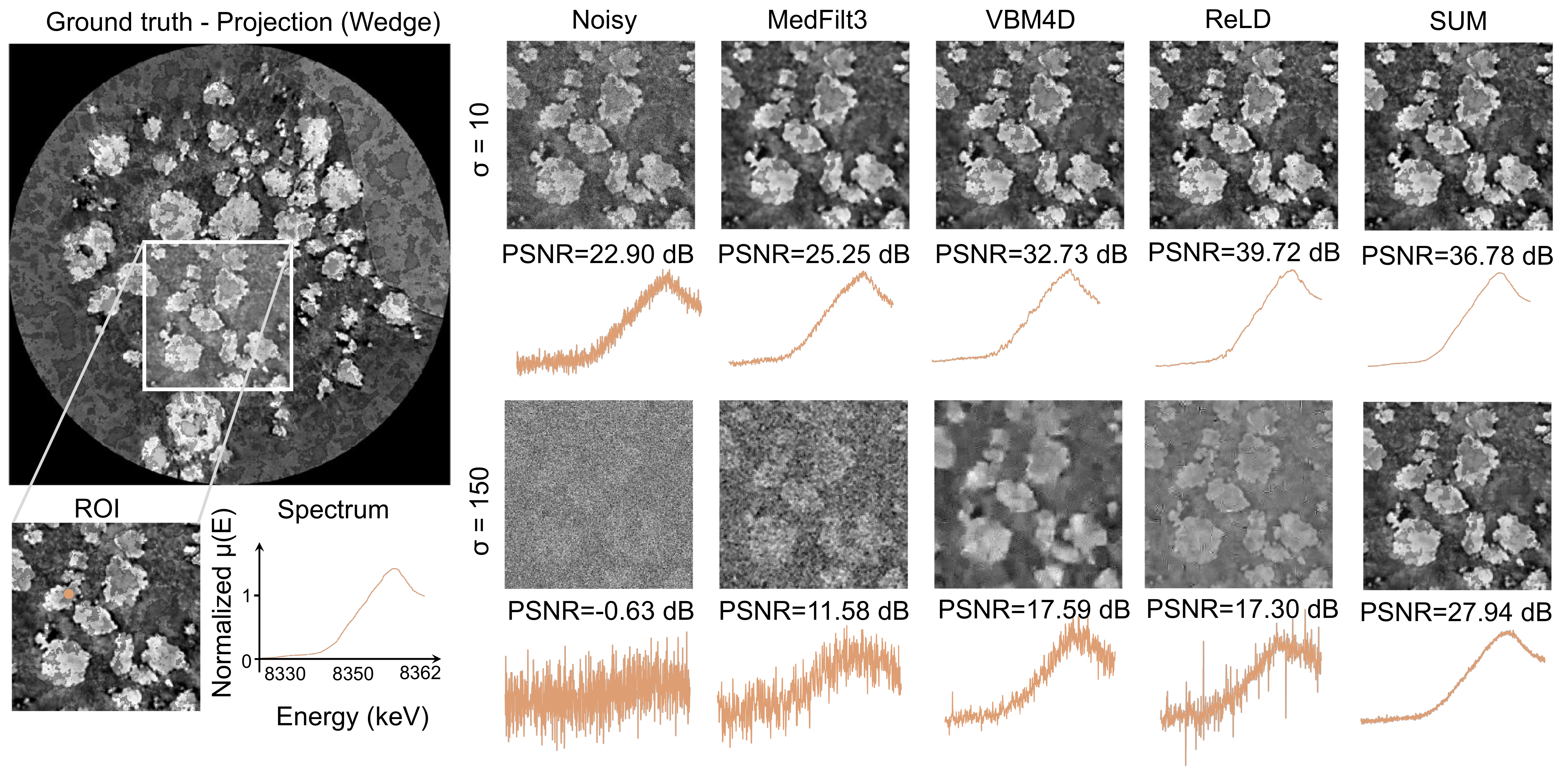}
    \vspace{-15pt}
    \caption{\textbf{Visual comparison of various methods under different noise conditions.} The PSNR values are calculated between the denoised and ground truth ROI (left). The spectra of the randomly selected pixel on the images (orange dot) are also plotted. The proposed algorithm SUM produces very good projection images with no noticeable artifacts even in high noise condition. The corresponding spectra have finer details than other approaches.}
    \label{fig7}
    \vspace{-10pt}
\end{figure*} 
\subsection{Analysis of the proposed algorithm}
\label{analysis}
\textbf{Selection of the Off-the-shelf Gaussian Denoiser.} The flexibility of the proposed algorithm allows us to plug-in any kinds of state-of-the-art \textit{image} denoising approaches. Here we focus on two representative algorithms: BM3D~\cite{dabov2007image} and DnCNN~\cite{zhang2017beyond}. As shown in~\cref{fig4}, both methods can significantly improve the PSNRs and correlations under different conditions, while SUM with BM3D obtains slightly better results. However, the running time of SUM with DnCNN is much smaller than the other, especially when the data size is large. The part of reasons is that DnCNN benefits from GPU computation while the implementation of BM3D is on CPU only. We adopt the DnCNN for the following experiments. Note that naively applying these methods to individual frame is undesirable (\textbf{Suppl. B}).

\textbf{Effect on the data sampling-rate.} Even if the proposed denoising approach is initially aiming at reducing the exposure time, it turns out that it also helps minimize the number of energy points for a reasonably good chemical map. As shown in~\cref{fig5}, we apply the SUM algorithm to the reduced-sampling-rate data and calculate the correlation with the ground truth under different noise conditions. The energy points are randomly selected. The results show that in order to get over 0.8 correlation, only 10\% data is needed if the noise is small, and 40\% data is needed if the noise is large, with the help of denoising.

\begin{table}[!ht]
    \centering
    \footnotesize{
    \begin{tabular}{ccccccc}
        \toprule
        
        \multicolumn{1}{l}{\phantom} &
        \multicolumn{3}{c}{$\sigma = 10$}    &
        \multicolumn{3}{c}{$\sigma = 150$}   \\
        \cmidrule(lr){2-4}
        \cmidrule(lr){5-7}
        &
        \multicolumn{1}{c}{Small} &
        \multicolumn{1}{c}{Medium} &
        \multicolumn{1}{c}{Large} &
        \multicolumn{1}{c}{Small} &
        \multicolumn{1}{c}{Medium} &
        \multicolumn{1}{c}{Large} \\

        \midrule
        Noisy & 0.63 & 0.58 & 0.54 & 0.04 & 0.04 & 0.04    \\
         MedFilt3 & 0.69 & 0.65 & 0.67 & 0.07 & 0.04 & 0.04    \\
         SVD & 0.77 & 0.71& 0.67& 0.31 & 0.31 & 0.22   \\
          ReLD & 0.79 & \textbf{0.75}& 0.72& 0.06 & 0.08 & 0.07   \\
        VBM4D & 0.77 & 0.71& 0.70& 0.15 & 0.17 & 0.16     \\
       
        SUM  & \textbf{0.84} & \textbf{0.78}& \textbf{0.78} & \textbf{0.76}& \textbf{0.72} & \textbf{0.72}   \\
        \bottomrule 
    \end{tabular}
    }
    \vspace{-5pt}
    \caption{\textbf{Demonstration of robustness to jitter motion.} The correlation coefficients between the ground truth and calculated chemical maps obtained from various algorithms under different noise levels ($\sigma=10$ and 150) and jitter amplitudes (small, medium and large). The best results within a 0.3 margin are shown in boldface.
        } 
    \label{tab:jittermotion}
    \vspace{-15pt}
\end{table}
\begin{figure*}
  \centering
   
    \includegraphics[scale=0.26]{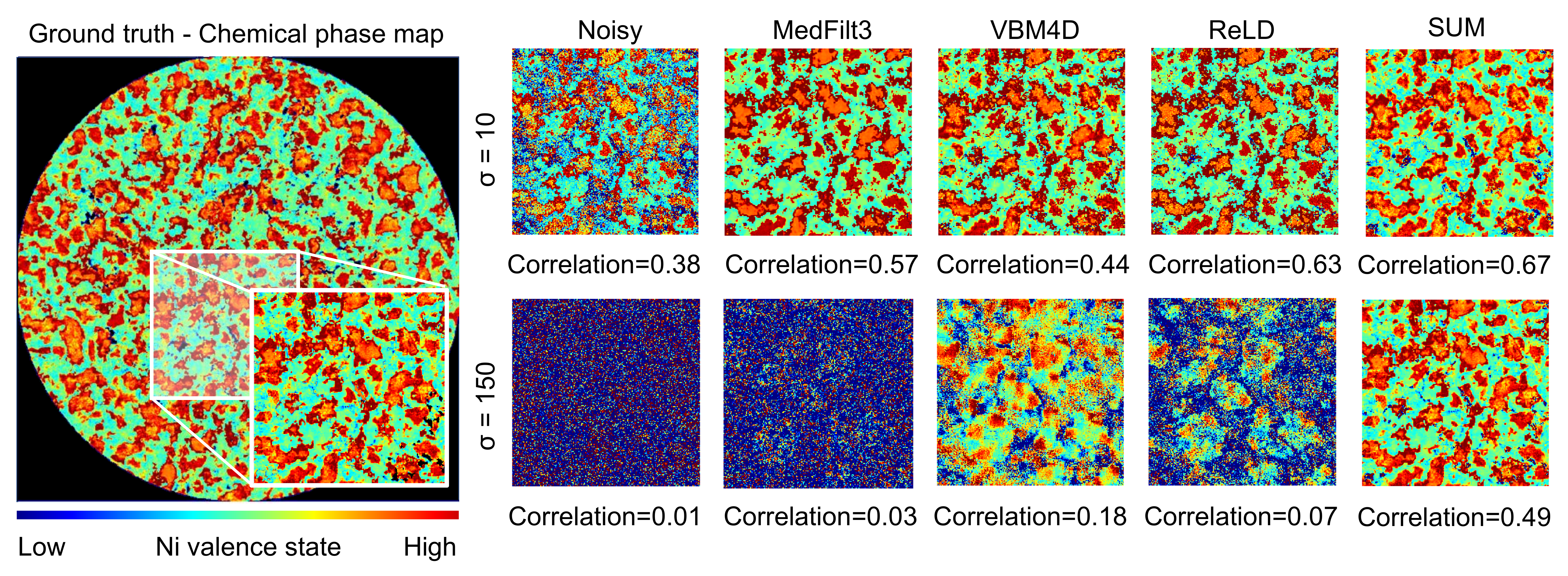}
    \vspace{-5pt}
    \caption{\textbf{Comparison of the chemical phase maps under different noise conditions.} The correlation coefficients to the ROI of the ground truth chemical phase map (left) are shown below the maps obtained from different algorithms. The proposed algorithm SUM produces highly accurate chemical phase maps.}
    \label{fig8}
    \vspace{-10pt}
\end{figure*} 
\vspace{-4pt}
\subsection{Comparison with others}
\vspace{-4pt}
As benchmarks for comparisons, we evaluate our method against six denoising techniques. They are categorized into three groups: Baseline (MedFilt3 - 3D median filtering, SVD), Traditional (ReLD~\cite{guo2014online}, VBM4D~\cite{maggioni2012video}), and Unsupervised CNN (UDVD~\cite{sheth2021unsupervised}, RFR~\cite{lee2021restore}). The optimal thresholding value of SVD is determined through the similar procedure as in~\cref{method}. Supervised CNN approaches such as FastDVDnet~\cite{tassano2020fastdvdnet} are not included here since the training dataset is limited and the pre-trained model on natural videos performs poorly in our case. 
\begin{table}[!t]
    \centering
    \vspace{-2pt}
    \footnotesize{
    \begin{tabular}{ccccccc}
        \toprule
        
        \multicolumn{1}{c}{\phantom} &
        \multicolumn{1}{c}{MedFilt3} &
        \multicolumn{1}{c}{SVD} &
        \multicolumn{1}{c}{ReLD} &
        \multicolumn{1}{c}{VBM4D} &
        \multicolumn{1}{c}{SUM}  \\

        \midrule
         Particle & \textbf{1.48} & 6.97 & 227.39 & 650.26  & 3.36   \\
        \midrule
        Wedge & 15.47 & \textbf{11.88} & 1620.44& 4196.10  & 14.40   \\
        
        \bottomrule 
    \end{tabular}
    }
    \vspace{-5pt}
    \caption{\textbf{Comparison of the computational time (units: seconds)}. Particle and Wedge have the image size of $379 \times 520 \times 969$  and $1193\times 1193\times 969$, respectively.
        }
    \label{tab:runningtime}
    \vspace{-13pt}
\end{table}

\textbf{Different Noise Levels.} \Cref{tab:test_data} reports the FPSNR and correlation results of two typical images we have obtained from various algorithms. The best results are shown in boldface. More results can be found in \textbf{Suppl. C}. We observe that SVD achieves the best results for the condition $\sigma=10$. This is not entirely surprising since SVD provides the optimal energy compaction in the least square sense. When the noise is small, SVD can effectively filtering out noise. As the noise level increases, SVD fails to preserve useful temporal information thus the performance drops significantly. The proposed SUM algorithm consistently achieves better results than other approaches and is very robust to a wide range of noise levels. In particular, significant improvements are observed at $\sigma=150$, where the chemical map is still well reconstructed.  

\Cref{fig7,fig8} show the visual comparison of the projection images and the chemical maps from various approaches under different noise conditions, respectively. We observe that our method preserves various image details, while introducing very few artifacts. The spectrum of a single pixel matches the ground truth. The chemical phase map obtained from the proposed algorithm has also the highest correlation with the ground truth. It is worth mentioning that when the noise level is high, it is almost impossible to retrieve the image from a single noisy frame. However, the spectrum information is spanned across multiple frames, which makes the restoration possible by the proposed approach.

\begin{figure}
  \centering
   \vspace{-5pt}
    \includegraphics[scale=0.26]{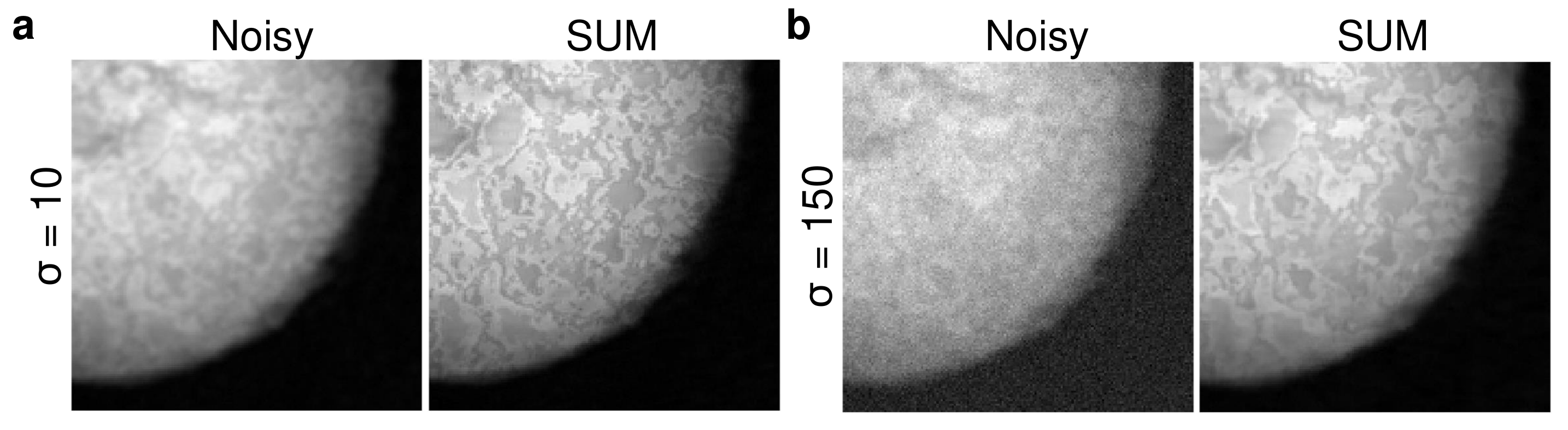}
    \vspace{-5pt}
    \caption{\textbf{SUM helps jitter correction.} The averaged images of the motion-corrected noisy input and denoised output by SUM across energies under low noise (a) and high noise (b) conditions, respectively. Structural details are clearly preserved after denoising, which indicates the image sequences are well-aligned.}
    \vspace{-15pt}
    \label{fig9}
\end{figure} 

\textbf{Motion Robustness.} It should be note that, because zone plate focusing is energy dependent, zone plate refocusing is performed for each energy and may result in jitter motion owing to poor motor precision. To evaluate the robustness of denoising algorithms to jitter motion, we simulate the data with random motion at different levels (\textit{Small}, \textit{Medium} and \textit{Large}), which are corresponding to jitter amplitudes 2, 4 and 6 pixels respectively. The motion correction is performed by NoRMCorre~\cite{pnevmatikakis2017normcorre} before calculating the chemical phase map. As shown in~\cref{tab:jittermotion}, the SUM algorithm consistently outperforms other approaches often by a significant margin, which demonstrates the robustness of the proposed denoising approach. 

Moreover, as shown in~\cref{fig9}, compared with the motion correction on raw noisy data, the denoising helps correct the jitter motion as well. The averaged images after denoising show clearer structural details. This is another important application of the proposed denoising algorithm for X-ray imaging in practice.

\textbf{Running Time.} \Cref{tab:runningtime} reports the comparison of computational time. It can be seen that our method is substantially faster than ReLD and VBM4D, and comparable with the simple approaches MedFilt3 and SVD. As observed, the proposed SUM algorithm is roughly 200 and 300 times faster than VBM4D for small and large images, respectively. It is worth mentioning that SUM is by nature highly parallelizable for even faster processing because each coefficient can be processed independently of the others.

\subsection{Results of real dataset}

We now apply the SUM algorithm to the restoration of real TXM-XANES data, which is the image of multiple NCM particles from a charged cathode. The exposure time for single frame is 0.5 seconds and data from in total 117 energy points spanning from 8180 eV to 8562 eV are recorded.   

\begin{figure}
  \centering
   
    \includegraphics[scale=0.37]{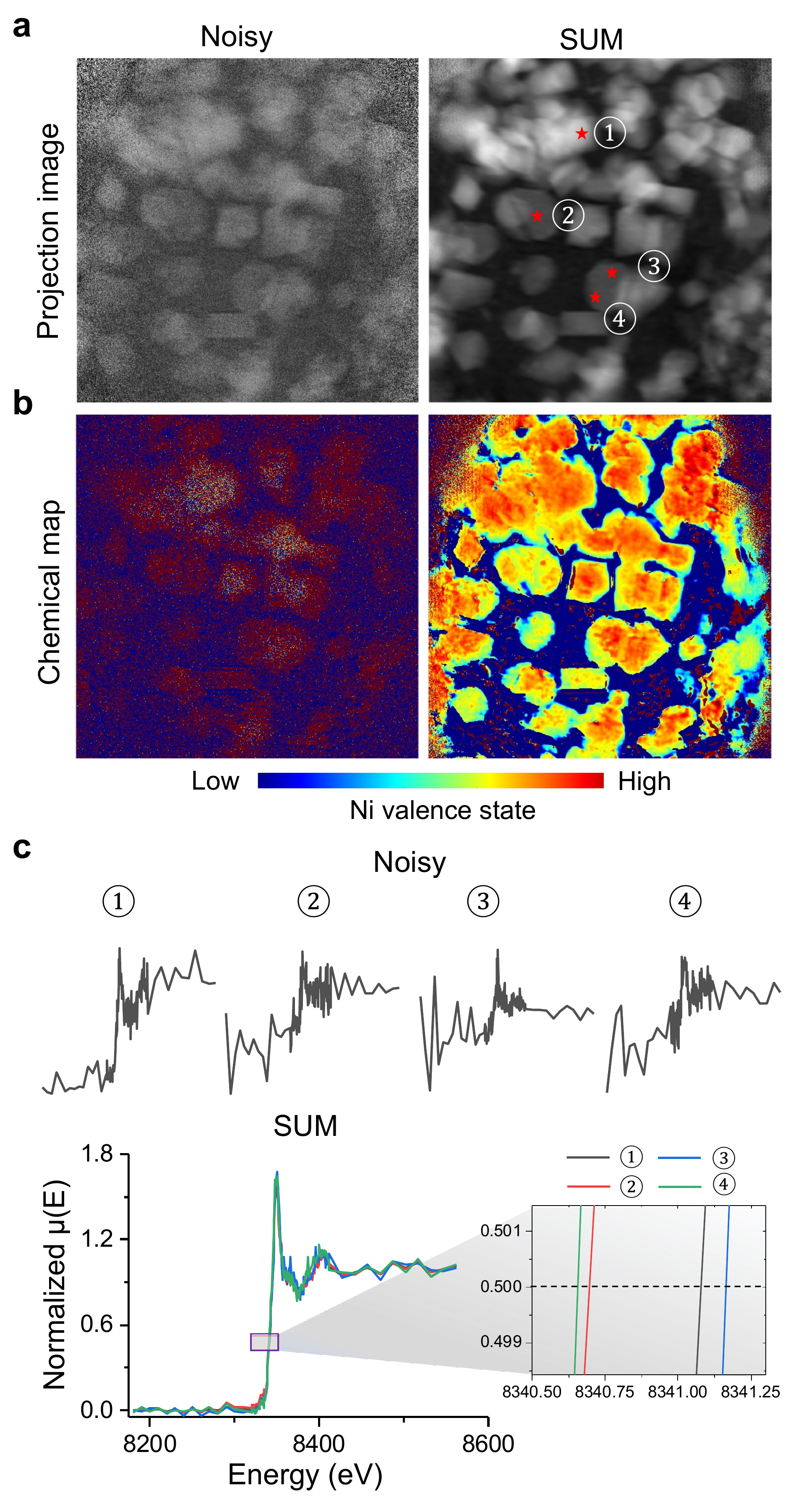}
    \vspace{-10pt}
    \caption{\textbf{SUM improves the interpretability of real data.} (a) Low SNR projection image of TXM-XANES recording (left) and its corresponding denoised image by SUM (right). (b) The fitted chemical state maps from the noisy and denoised data, respectively. Colors represent the level of Ni valence state. (c) Spectra of selected four single pixels in (a) from the noisy (top) and denoised data (bottom), respectively. The region of edge energy position (0.5 edge position) is amplified for better illustration. The inter- and intra- particle state differences are revealed with the help of SUM.}
    \vspace{-15pt}
    \label{fig10}
\end{figure}

As shown in~\cref{fig10}a, each projection image is very noisy and the resulting chemical map does not contain meaningful information. After applying the proposed SUM algorithm to the raw data, we can clearly see the morphological structure of NCM particles. In addition, the chemical map from the denoised data shows the inhomogeneous reactions of battery electrodes. The Ni valence states of some NCM particles are relatively high while those of others are low. The denoising by SUM provides the possibilities of deeper insights into spatiotemporally electrochemical reactions and ultimately helps optimize the design of the composite electrodes.

We plot the spectra of four single pixels at different locations before and after denoising in~\cref{fig10}c. The raw data is too noisy to be analysed reliably. With the help of SUM, the transition of the chemical states at four different locations is revealed with high fidelity, which shows the inter- and intra- particle differences during battery charging and discharging process.     

\vspace{-2pt}
\section{Limitations}
\vspace{-2pt}

\textbf{Memory Issue.} The standard SVD calculation in the proposed approach may be not efficient for extremely large matrix (\eg full resolution TXM images and thousands of energy points) that cannot fit into the computer's main memory. This limitation can be addressed by applying randomized numerical linear algebra~\cite{drineas2016randnla,martinsson2011randomized}.

\textbf{Noise Assumption.} We assume the noise distribution of X-ray images is Gaussian and additive in~\cref{model}. Though well approximated by Gaussian noise in many cases, more complicated noise types, such as mixed Poisson-Gaussian noise, should be considered in the extremely low-dosage situations. The variance stabilization transformation~\cite{makitalo2010optimal} or a new unbiased estimate considering the noise statistics~\cite{luisier2010image} could be vital to addressing this limitation.   

\vspace{-2pt}
\section{Conclusion}
\vspace{-2pt}
We propose a new denoising algorithm for the fundamental and widely-used X-ray imaging technique. The proposed SUM approach outperforms related state-of-the-art techniques, both qualitatively and computationally. The effectiveness and low computational cost of the SUM algorithm offers the possibilities of fast and high-sensitivity chemical imaging. One of the major advantages is to identify unknown chemical phases with high precision in the sample since this is only possible if an accurate XANES spectrum is provided for each individual pixel. The proposed approach can be easily extended to 3D chemical imaging by rotating the sample~\cite{kuppan2017phase}. Another direction of our future work is to adaptively select the energy points to further reduce the acquisition time and enable efficient autonomous experiments~\cite{noack2021gaussian}. Moreover, the proposed algorithm is not limited to X-ray imaging. Multispectral image restoration could be another suitable application. 

{\small
\bibliographystyle{ieee_fullname}
\bibliography{main}
}

\end{document}